\documentclass[preprint,aps,showpacs,nofootinbib,preprintnumbers,amsmath,amssymb]{revtex4-1}
\usepackage{epsfig}
\usepackage{bm}
\usepackage[usenames ,dvipsnames]{xcolor}
\usepackage{slashed}
\usepackage{caption}
\usepackage{subcaption}
\usepackage{graphicx,color}

\usepackage{multirow}

\begin{document}
	\title{Sizable Time-reversal  Violating Effects in  Bottom Baryon Decays}
		\author{Chia-Wei Liu and Chao-Qiang Geng}
	\affiliation{
		School of Fundamental Physics and Mathematical Sciences, Hangzhou Institute for Advanced Study, UCAS, Hangzhou 310024, China 
	}\date{\today}

\begin{abstract}
We study  time-reversal~(T) asymmetries in the charmless two-body decays of  antitriplet bottom baryons. We find that in ${\bf B}_b \to {\bf B}_n P$ with ${\bf B}_b = (\Xi _b ^- , \,\Xi_b^0, \,\Lambda_b)$ and ${\bf B}_n(P)$  a octet baryon (pseudoscalar meson), the positive and negative helicity amplitudes are both sizable, resulting in  the large T-violating effects. Particularly,  the T-odd parameters, $\beta_w$, for the color-enhanced channels are expected to be around $-40\%$ and $25\%$ in  $b\to s $ and $b\to d $ transitions for the standard model, which can be measured by the experiments at LHCb, respectively. On the other hand,  in ${\bf B}_b \to {\bf B}_n V(\gamma)$ with  $V(\gamma)$ a vector meson~(photon), the decays are predominated by a single helicity amplitude, and the T violating effects by the correlations among spins are suppressed.   We also explore  the T violating observables  for  the $\Xi_b$ decays based on the angular distributions.
\end{abstract}

\maketitle

\section{Introduction}
It is well known that  time-reversal~(T) and CP violations are equivalent due to the CPT theorem~\cite{CPT}. The indirect and direct evidences of  
CP violation (CPV) were found in 1964~\cite{Christenson:1964fg} and 2001~\cite{BaBar:2001ags,Belle:2001zzw}, respectively. 
On the other hand, the first direct evidence on T violation (TV) was long-awaited until 2012 via the $B^0 - \overline{B}^0$ oscillation~\cite{BaBar:2012bwc}. 
As  baryon-antibaryon oscillations are forbidden, a direct measurement on TV in baryon decays is still lacking. Nevertheless, we can study TV with the triple-vector product asymmetries~(TPAs), which can be viewed as indirect evidences on TV.
The TPAs have been intensively studied in the mesonic weak decays~\cite{ref1,ref4,ref7,ref5,ref6,ref8,Li:2010ra,Geng:BtoVV,Datta:2003mj,Kramer:1993yu,Valencia:1988it,Chiang:1999qn,London:2004ws,LondonBtoVT,Gronau:2011cf,Ali:1998gb,Kagan:2004uw}. 
In terms of angular analyses, there are over 40 experimental values related to TPAs in $B\to VV'$ with $V^{(\prime)}= (\rho,\omega,K^*,\phi)$~\cite{LHCb:2012mid,pdg,LHCb:2014xzf,LHCb:2018hsm,LHCb:2019jgw}. 
However, there is still no experimental evidence on a TPA in the bottom baryon decays.
 Our goal in this work is to seek  some measurable  large 
T violating effects (TVEs) in the bottom baryon decays.

Compared to the spinless  mesons, the antitriplet bottom baryons~(${\bf B}_b$) are polarized, providing a possibility for observing TPAs  
even in the two body decays of ${\bf B}_b \to {\bf B}_n (P,V,\gamma)$ with   ${\bf B}_n$, $P(V)$ and $\gamma$ being
the octet baryon, pseudoscalar (vector) meson and photon, respectively.
Recently, the lifetimes of ${\bf B}_b$  have been updated with high precision at LHCb~\cite{XibTau,XibmTau,CMS:2017ygm,pdg}, and CP violating effects have been studied~\cite{XibBr,XibCP,LbPK}.
In particular, a full angular analysis has been carried out in $\Lambda_b \to J/\psi \Lambda$~\cite{ExpPolarized1,ExpPolarized2,ExpPolarized3}.
Clearly, the experimental progresses on ${\bf B}_b$ result in a great
 opportunity for us to explore TV in the two-body decays of ${\bf B}_b$.
On the theoretical side for ${\bf B}_b$ decays, many efforts have been made~\cite{Dery:2020lbc,Franklin:2020bvb,Han:2021gkl,Roy:2019cky,Roy:2020nyx,GeneralGeng:2016kjv,GeneralGeng:2021nkl,GeneralHsiao:2017tif,GeneralModifiedBagModel,NaiveFacZhao:2018zcb,NaiveKhodjamirian:2011jp,KornerSM,KornerSM2,Boer:2019zmp,Zhu:2018jet,Lu:2009cm}, but only few of them have discussed TVEs~\cite{ref2,ref3,Lambdab to LambdaV,French0,French1,French2,Gronau:2015gha}.
For the weak transitions  described by the effective Hamiltonian~(${\cal H}_{eff}$)~\cite{Buras:1991jm}, 
based on the Hermitian nature of ${\cal H}_{eff}$ and  T-symmetry, we  have the complex phase approach with
\begin{equation}\label{complexphase}
H_\lambda =  \langle \lambda ; \text{``out''}  |{\cal H}_{eff} | {\bf B}_b \rangle = \langle {\bf B}_b | {\cal H}_{eff} | \lambda; \text{``in''} \rangle  \,,
\end{equation}
where  $\lambda$ are the helicities of the final states. By ignoring  the final state interactions~(FSIs), 
``in'' and ``out'' are interchangable, such that Eq.~\eqref{complexphase} demands  $H_\lambda$ to be real
due to the anti-liner property of the T-transformation. This approach  is a common method in exploring  TV, 
which has been shown closely related to the TPAs involving momentum and spin~\cite{ref2,ref3,French0,Lambdab to LambdaV,Gronau:2015gha,French1,French2}. 
However, there are some defects in the complex phase approach. The relation between phases and TV is not intuitively understandable. In addition, 
the analyses of the TPAs carried out in the  literature suffer a theoretical inconsistency~\cite{Lambdab to LambdaV}.

In this study, we use a different approach
to define an asymmetry parameter with a T-odd scalar operator $\hat{T}$ in the ${\bf B}_b$ decays,  given by
\begin{equation}
\Delta_{\hat{T}} \equiv \frac{\Gamma(\lambda_t
>0) - \Gamma(\lambda_t
<0) }{\Gamma(\lambda_t
>0) +\Gamma(\lambda_t
<0) }\,,
\end{equation}
where $\lambda_t$ are the eigenvalues of $\hat{T}$. 
If ${\cal H}_{eff}$ respects T-symmetry, $\pm \lambda_t $  are related by
\begin{equation}\label{T-transform}
|\langle \lambda_t ; \text{``out''} | {\cal H}_{eff} | {\bf B}_b\rangle|^2 - |\langle {\bf B}_b | {\cal H}_{eff} | -\lambda_t ; \text{``in''}\rangle|^2=0\,.
\end{equation}

 Note that  if ``in'' and ``out''  are interchangable, the left-handed side of Eq.~\eqref{T-transform} is proportional to the numerator in $\Delta_{\hat{T}} $,
 leading to that a nonzero value of $\Delta_{\hat{T}}$ would be a signal of TV. Nonetheless,
 under FSIs, we have 
 \begin{equation}
\langle \mp \lambda_t ; \text{``out''}  | \pm \lambda_t ; \text{``in''}\rangle \neq 0\,, 
 \end{equation}
 describing the rescattering effects between $\pm \lambda_t$.
 In this case,  $\Delta_{\hat{T}} $ is referred to as a naive T-odd observable.  
 To subtract the effects of FSIs, we utilize the CPT symmetry and thereby define a true T violating observable, which is also a CP violating one, as~\cite{Gronau:2011cf,Gronau:2015gha}
 \begin{equation}\label{True}
{\cal T} \equiv (\Delta _{\hat{T}}  - (\pm ) \overline{\Delta}_{\hat{T}} )/2 \,,
\end{equation}
where the overline denotes the charge conjugate and the sign corresponds to the parity of ${\hat T}$.  
The most simple cases for $\hat{T}$ are the triple  products of the  spins~($\vec{s}$) and 3-momenta~($\vec{p}$) of the particles in the decays.
Since there is only one independent 3-momentum in the two-body decays of ${\bf B}_b \to {\bf B}_n (P,V,\gamma)$,  
$\hat{T}$ contains at least two spins,  called as the spin correlations. 
Consequently, to observe a TVE, a decay must involve at least two different helicities. 
In this work, we explore  TV by concentrating on the T violating observables in ${\bf B}_b \to {\bf B}_n M$.

This paper is organized as follows. In Sec.~\MakeUppercase{\romannumeral 2}, we discuss TV in the two-body decays
of ${\bf B}_b \to {\bf B}_n (P,V,\gamma)$. 
We also provide the numerical results in the standard model for ${\bf B}_b\to {\bf B}_n P$.
In Sec.~\MakeUppercase{\romannumeral 3}, we study the angular distributions of the ${\bf B}_b$ decays. We give our conclusions   
in Sec.~\MakeUppercase{\romannumeral 4}.
	
\section{T-odd observables}
For the ${\bf B}_b$ decays, 
 the effective Hamiltonian  is given by~\cite{Buras:1991jm}
\begin{eqnarray}
&&{\cal H}_{eff} = \frac{G_F}{\sqrt{2}}
\sum_{f=d,s}\left[
V_{ub}V_{uf}^* \left(
C_1O^f_1 + C_2 O^f_2
\right) - V_{tb}V_{tq}^* \sum_{i=3}^{10}C_i O^f_i
\right]\,,
\end{eqnarray}
with  
\begin{eqnarray}
&&O_1^f =  (\overline{u}_\alpha \gamma^\mu L  b_\alpha) (\overline{f}_\beta \gamma_\mu L u_\beta) \,,\,\,\,\,~~~~~~~~~~O_2^f =(\overline{u}_\alpha \gamma^\mu L b_\beta) (\overline{f}_\beta \gamma_\mu L u_\alpha)\,,\nonumber\\ 
&&O_3^f = (\overline{f}_\alpha \gamma^\mu L b_\alpha) \sum_{q} (\overline{q}_\beta  \gamma_\mu L q_\beta)\,,\,\,\,\,~~~~~O_4^f = (\overline{f}_\alpha \gamma^\mu L b_\beta) \sum_{q}(\overline{q}_\beta \gamma_\mu L q_\alpha)\,,\nonumber\\
&&O_5^f = (\overline{f}_\alpha \gamma^\mu L b_\alpha) \sum_{q}(\overline{q}_\beta \gamma_\mu R q_\beta)\,,\,\,\,\,~~~~~O_6^f = (\overline{f}_\alpha \gamma^\mu L  b_\beta) \sum_{q}(\overline{q}_\beta \gamma_\mu R q_\alpha)\,,\nonumber\\
&&O_7^f = \frac{3}{2}(\overline{f}_\alpha \gamma^\mu L b_\alpha) \sum_{q}e_{q}(\overline{q}_\beta \gamma_\mu R q_\beta)\,,\,\,\,\,O_8^f =\frac{3}{2} (\overline{f}_\alpha  \gamma^\mu L  b_\beta) \sum_{q}e_{q}(\overline{q}_\beta \gamma_\mu R q_\alpha)\,,\nonumber\\
&&O_9^f = \frac{3}{2}(\overline{f}_\alpha  \gamma^\mu L b_\alpha) \sum_{q}e_{q} (\overline{q}_\beta \gamma_\mu L  q_\beta)\,,\,\,\,\,O_{10}^f =\frac{3}{2} (\overline{f}_\alpha \gamma^\mu L  b_\beta) \sum_{q}e_{q}(\overline{q}_\beta \gamma_\mu L  q_\alpha)\,,\nonumber\\
&&O^f_{7\gamma} = \frac{e}{8 \pi^{2}} m_{\mathrm{b}} \bar{f}_{\alpha} \sigma^{\mu \nu}  R b_{\alpha } F_{\mu \nu}\,,\,\,\,\,O_{8 G}=\frac{g}{8 \pi^{2}} m_{\mathrm{b}} \bar{f}_{\alpha} \sigma^{\mu \nu} T_{\alpha \beta}^{a}  R b_{\beta} G_{\mu \nu}^{a}\,,
\end{eqnarray}
where $O_{1,2}$, $O_{3,4,5,6}$, $O_{7,8,9,10}$ and $O_{7\gamma,8G}$ correspond to  tree, penguin, electroweak-penguin and magnetic-penguin operators, respectively, $G_F$ is the Fermi constant, $V$ represents the CKM quark mixing matrix,
$\alpha$ and $\beta$ denote the color, $q=(u,d,s)$, and $L ~(R)$ stands for the chiral structure of $1\mp \gamma_5 $.

  In the radiative decays of ${\bf B}_b \to {\bf B}_n \gamma$, the relevant operator is $O^f_{7\gamma}$, in which $f$ is essentially left handed.
As the decays are predominated by a single helicity state of $\lambda_n=-1/2$,  the T-violating effects are expected to be suppressed.
In the factorization approach for ${\bf B}_b \to {\bf B}_n (P,V)$, $O_{1,2,3,4, 9,10}$  can only produce left-handed ${\bf B}_n$
since they involve only left-handed currents.
On the other hand, 
for $O_{5,6,7,8}$, based on the Fierz transformation,
one has that
 \begin{equation}
 (\overline{f} \gamma^\mu L b) \sum_{q}(\overline{q} \gamma_\mu R q) =  -2 (\overline{q} L b ) (\overline{f} R q) \,,
 \end{equation} 
 in which the chiral structure of $(\overline{q} L b)$  leads to  a right-handed ${\bf B}_n$, given by
\begin{equation}
\langle {\bf B}_n | \overline{q}L b |{\bf B}_b\rangle = \overline{u}_n (f_s - g_a\gamma_5) u_b\,,
\end{equation}
where $u_{b(n)}$ represents the Dirac spinor for ${\bf B}_{b(n)}$, and   $f_s(g_a)$ corresponds to the (pseudo)scalar form factor
with $f_s = g_a$ in the heavy quark limit.
Here,  the color index has not been explicitly shown. 
Clearly, the amplitudes between  $\lambda_n =-1/2$ and $\lambda_n =1/2$ from
	$O_{1,2,3,4,9,10}$ and $O_{5,6,7,8}$, respectively, can potentially have  interferences.
Furthermore, in ${\bf B}_b \to {\bf B}_n V$, the vector meson matrix elements of  (pseudo)scalar operators vanish, read as 
\begin{equation}
\langle V | \overline{f} R q | 0 \rangle = 0\,,
\end{equation}
resulting in that  the TVEs  are suppressed because the decays are dominated by a single helicity state~\cite{Lambdab to LambdaV}.
In contrast, in ${\bf B}_b \to {\bf B}_n P$, the pseudoscalar meson matrix elements are enhanced by the smallness of the light quark masses, given by 
\begin{equation}\label{enhancement}
 i \langle P | \overline{f} R q  | 0 \rangle = \frac{m_P^2}{m_f + m_q } f_P\,,
\end{equation}
with $m$ standing for the mass and $f_P$ the meson decay constants.
Since the denominators of the right side of Eq.~(\ref{enhancement})
are made of the light quark masses, $O_{5,6}$
 cause a  significant contribution to $\lambda_n = 1/2$, providing a  interference with $\lambda_n = -1/2$ from $O_{1,2}$.
We conclude that the decays of  ${\bf B}_b\to {\bf B}_n P$  are the most promising modes to  observe TVEs.

\begin{figure}
		\captionsetup{justification=raggedright,
		singlelinecheck=false
	}
	\includegraphics[width=.3\linewidth]{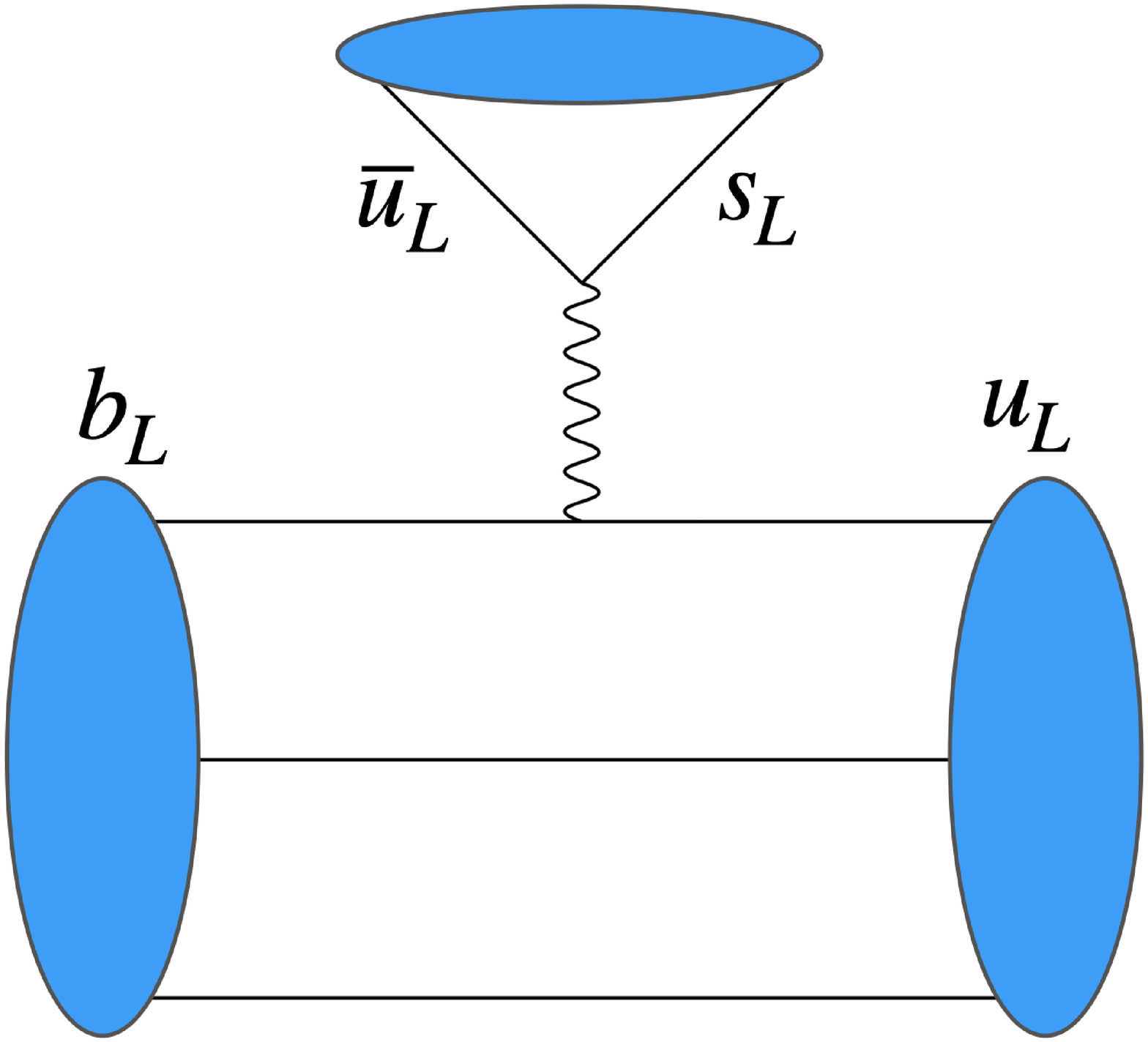}  
	~~~~~~~~~~
	\includegraphics[width=.3\linewidth]{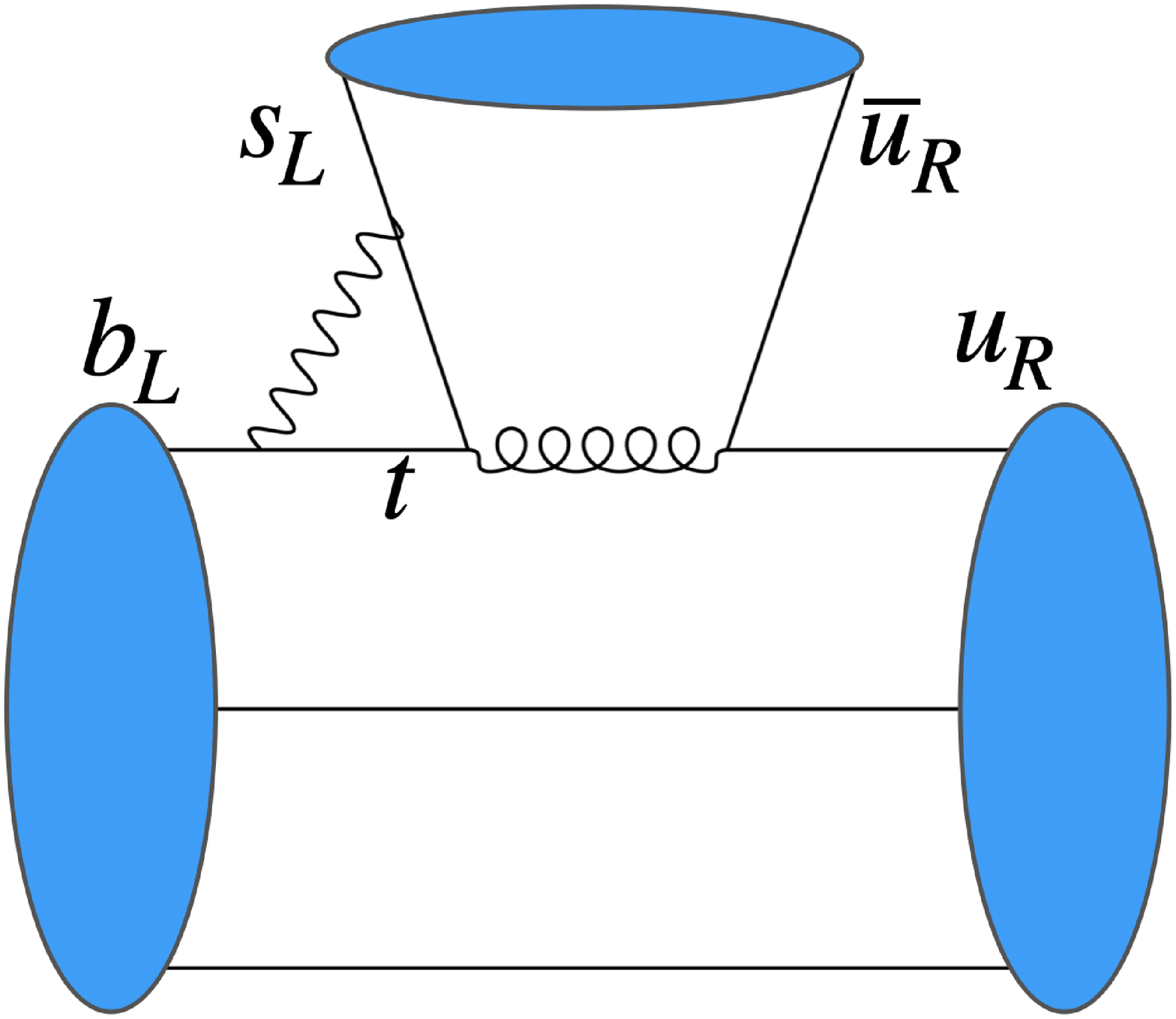}  
	\caption{
Topological diagrams for the color enhanced processes in ${\bf B}_b\to {\bf B}_n P$, where the left and right figures are associated with
 the CKM coefficients of $V_{ub}V_{us}^*$ and $V_{tb}V_{ts}^*$, respectively.
	}
\end{figure}

In  ${\bf B}_b\to {\bf B}_n P$, as $P$ are spinless,  the spin correlations only happen between ${\bf B}_b$ and ${\bf B}_n$. The most simple T-odd observables 
are  then described by
\begin{equation}\label{T3}
\hat{T} \equiv  (\hat{p} \times \vec{s} ) \cdot \vec{J}\,,
\end{equation}
where $\hat{ p}~(\vec{ s})$ are the 3-momenta~(spins) of ${\bf B}_n$, and $\vec{ J}$ are the angular momenta in  the final states. Since $\vec{ J}$ are always conserved, one can identify them as the spins of ${\bf B}_b$.
The  eigenstates of $\hat{T}$  are related to the helicity ones as~\cite{Lambdab to LambdaV}
\begin{equation}
|\lambda_{t}= \pm \frac{1}{2}\rangle = \frac{1}{\sqrt{2}}\left(
|H_+\rangle \pm i | H_-\rangle 
\right)\,,
\end{equation}
with $J = 1/2$ in both sides.
The naive T-odd observable is then given by the asymmetry between $\lambda_t= \pm 1/2$, read as~\cite{HXG}
\begin{equation}\label{beta}
\beta \equiv \frac{\Gamma(\lambda_t
	=\frac{1}{2}) - \Gamma(\lambda_t
	=-\frac{1}{2}) }{\Gamma(\lambda_t
	=\frac{1}{2}) +\Gamma(\lambda_t
	=-\frac{1}{2}) } =  \frac{2\Im(H_+H_-^*)}{|H_+|^2+|H_-|^2}=\frac{2\Im({\cal P}{\cal S}^*)}{|{\cal S}|^2+|{\cal P}|^2}\,,\end{equation}
where ${\cal S}$ and ${\cal P}$ are the amplitudes with negative and positive parities, respectively, and $H_\pm$ stand for the amplitudes with $\lambda_n= \pm 1/2$. 
Thereby, from Eq.~\eqref{True}, we can define the true and fake T-odd observables as~\cite{Gronau:2015gha}
\begin{equation}\label{TrueT3}
\beta_w = (\beta + \overline{\beta})/2\,,\,\,\,\,\,\,\,\beta_s = (\beta - \overline{\beta})/2\,,
\end{equation}
 related by the CP-odd and even phases, respectively. 
 We emphasize that 
 a nonzero value of $\beta_w$ implies not only TV but also CPV.
  To have a complete study
in ${\bf B}_b\to {\bf B}_n P$,
 we also  define the direct CP and up-down asymmetries by~\cite{HXG} 
\begin{eqnarray}
A_{\text{CP}} & \equiv &\frac{\Gamma -\overline{\Gamma}}{\Gamma+ \overline{\Gamma}}\,,
\\
\alpha_b &\equiv& \frac{\Gamma(\lambda_{n}=\frac{1}{2})-\Gamma(\lambda_{n}=-\frac{1}{2})}{\Gamma(\lambda_{n}=\frac{1}{2})+\Gamma(\lambda_{n}=-\frac{1}{2})}= \frac{2\Re({\cal P}{\cal S}^*)}{|{\cal S}|^2 + |{\cal P}|^2}\,,
\end{eqnarray}
respectively.
The linear combinations of $\alpha_b$ and $\overline{\alpha}_b$ with the overline denoting the charge conjugate are given by
\begin{eqnarray}
\alpha_{\text{av}} \equiv (\alpha_b - \overline{\alpha}_b)/2 \,,\,\,\,\,\,\,\,\,\,\,\,\Delta \alpha = \frac{\alpha_b + \overline{\alpha}_b}{\alpha_b - \overline{\alpha}_b}\,,
\end{eqnarray}
in which $\Delta \alpha$ is a CP-odd quantity.

It is well known that to have nonzero values of $A_{\text{CP}}$  and $\Delta \alpha$, nonvanishing  strong phases are both required.
Explicitly, one has that
\begin{equation}
A_{\text{CP}} , \Delta \alpha \sim \sin \phi_s \sin \phi _w \,,
\end{equation}
where $\phi_s$ and $\phi_w$ correspond to  the CP-even and CP-odd phases between ${\cal S}$ and ${\cal P}$, respectively.
On the other hand, one can show that
\begin{equation}
\beta_w \propto \cos \phi_s \sin \phi_w \,,\,\,\,\,\,\,\beta_s \propto \cos \phi_w \sin \phi_s\,.
\end{equation}
Clearly, a nonzero strong phase is not necessary in   $\beta_w$. In fact, when $\phi_s\to 0$, $\beta_w$ becomes maximal, whereas 
$(A_{\text{CP}},\Delta \alpha)\to 0$.
Since the strong phases are much less understood compared to  the weak ones, $\beta_w$ could be  a more promising T and CP violating observable compared to the others.

In ${\bf B}_b \to {\bf B}_n P$,  TV depends   on the relative CP-odd phase between $V_{ub}V_{uf}^*$ and $V_{tb}V_{tf}^*$ in the different helicity amplitudes~(see Eq.~\eqref{beta} and FIG.~1), related to  the interferences between $O_{1,2}^f$ and $O_{5,6}^f$. Since $C_{1,6}$  are much larger compared to $C_{2,5}$, respectively, one expects that  large T-asymmetries can be induced  in the color-enhanced decays.

In practice,
	the decays are often dominated by one of the helicities.
	For $|H_\pm|^2/|H_\mp|^2\gg 0 $,
	we have the approximations, given by
	\begin{equation}\label{limit}
	\beta \approx 2 \Im\left(\frac{H_+H_-^*}{H_\pm H_\pm^*}\right) = \mp 2 \Im\left(
	\frac{H_\mp}{H_\pm}
	\right)\,.
	\end{equation} 
As we will see subsequently,	in ${\bf B}_b \to {\bf B}_n P$, the transitions associated with 
$b\to s$ and $b\to d $  are predominated by  $H_+$ and $H_-$, respectively. Using the estimations of
$H_-\propto c_1V_{ub}V_{uf}^*$ and $H_+ \propto -c_6 V_{tb}V_{tf}^*$, we have
	\begin{equation}\label{newadd1}
	\beta_w \approx 2\Im \left(
	\frac{V_{ub}V_{us}^*c_1}{V_{tb}V_{ts}^* c_6}
	\right) = \frac{2 c_1}{c_6}  \lambda^2 \eta \approx -0.7 \,,
	\end{equation}
	and 
	\begin{equation}\label{newadd2}
\beta_w \approx -2\Im \left(
\frac{V_{tb}V_{td}^*c_6}{V_{ub}V_{ud}^* c_1}
\right)\approx -\frac{2c_6}{c_1} \frac{\eta }{\rho^2 + \eta^2}\approx 0.2\,,
\end{equation}
for $b\to s $ and $b\to d$ transitions, respectively.	
Here, we have used $c_6/c_1=0.04$~\cite{Buras:1991jm} and the Wolfenstein parametrization for the CKM matrix, given by~\cite{Wolfenstein:1983yz}
\begin{equation}
\left[\begin{array}{lll}
V_{u d} & V_{u s} & V_{u b} \\
V_{c d} & V_{c s} & V_{c b} \\
V_{t d} & V_{t s} & V_{t b}
\end{array}\right] = 
\left[\begin{array}{ccc}
1-\lambda^{2} / 2 & \lambda & A \lambda^{3}(\rho-i \eta) \\
-\lambda & 1-\lambda^{2} / 2 & A \lambda^{2} \\
A \lambda^{3}(1-\rho-i \eta) & -A \lambda^{2} & 1
\end{array}\right] + O(\lambda^4) \,.
\end{equation}
It is interesting to point out that although $|V_{ub}V_{us}^*/V_{tb}V^*_{ts}|$  and  $|V_{ub}V_{ud}^*/V_{tb}V^*_{td}|$ are very different, $|\beta_w|$ are in the same size for $b\to s$ and $b\to d$ transitions. It is due to that they are applied at the different limits of Eq.~\eqref{limit}, allowing them to be both sizable.

It is known that the factorization approach works well in the color-enhanced channels~\cite{EffectiveWilson}.
To get the more precise results, we adopt the generalized factorization approach, in which the renormalization dependency in the naive factorization has been absorbed into the effective Wilson coefficients to the next to leading log precision~\cite{EffectiveWilson}. Accordingly, the color-enhanced amplitudes are  given as~\cite{GeneralModifiedBagModel,GeneralGeng:2021nkl,GeneralGeng:2016kjv,GeneralHsiao:2017tif}
\begin{equation}
\langle {\bf B}_n P | {\cal H}_{eff} | {\bf B}_b\rangle = i
 \overline{u}_n (A +B\gamma_5) u_b\,,
\end{equation}
with 
\begin{eqnarray}\label{calcula}
&&A= \frac{G_F}{\sqrt{2}} f_P\left\{ m_b V_{ub} V_{uf}^*a_1 -  V_{tb} V_{tf}^*\left[m_b (a_4 + a_{10} )   +  \frac{2m_P^2 }{m_u+m_f} (a_6+a_8) \right] \right\}f_s\,,\nonumber\\
&&B = \frac{G_F}{\sqrt{2}} f_P \left\{ m_b V_{ub} V_{uf}^*a_1 -  V_{tb} V_{tf}^*\left[m_b (a_4 + a_{10} )   - \frac{2m_P^2 }{m_u+m_f} (a_6+a_8) \right] \right\}g_a\,,
\end{eqnarray}
where  $a_i$ are the effective Wilson coefficients~\cite{GeneralGeng:2021nkl,GeneralGeng:2016kjv,GeneralHsiao:2017tif}.
The helicity amplitudes with definite angular momenta are then given as~\cite{Korner:1992wi}
\begin{equation}
H_\pm = \sqrt{Q_+} A \mp \sqrt{Q_-} B\,,
\end{equation}
with $Q_\pm = \sqrt{(m_{{\bf B}_b} \pm m_{{\bf B}_n})^2 - m_P^2}$, and the decay widths are  given by 
\begin{equation}
\Gamma = \frac{|\vec{p}_c|}{16\pi m_{{\bf B}_b^2} }\left(
	|H_+|^2 +| H_-|^2
	\right)\,,
\end{equation}
where $\vec{p}_c$ are the 3-momenta of ${\bf B}_n$ in the central mass frames.
At the heavy quark limit, we have that $Q_+ = Q_-$. It is straight forward to see that the terms associated with $(a_6+a_8)$ contribute to $H_+$,
 while the others $H_-$ only.

We evaluate
the form factors of $f_s$ and $g_a$ in Eq.~\eqref{calcula}
 from the modified bag model~\cite{GeneralModifiedBagModel} with the numerical results given in Table~\MakeUppercase{\romannumeral 1}~\cite{footnote1}, where the first and second uncertainties in the branching ratios come from the bag radius of $R = (4.8 \pm 0.2$ GeV)$^{-1}$~\cite{GeneralModifiedBagModel} and   CKM elements~\cite{pdg}, respectively.
\begin{table}\label{table}
\begin{center}
		\captionsetup{justification=raggedright,
		singlelinecheck=false
	}
	\caption{Numerical results in ${\bf B}_b \to {\bf B}_n P$, 
	where the first and second  uncertainties in the branching ratios come from the bag radius~\cite{GeneralModifiedBagModel} and  CKM elements~\cite{pdg}, respectively.
	 }
	\begin{tabular}{l c|ccccc }
		\hline
		\multicolumn{2}{c|}{\multirow{2}{*}{channels}} &$10^6{\cal B}$&$\alpha_{\text{av}}$&$\beta_s$&$\phi_s$\\
		&&$A_{\text{CP}}$&$\Delta \alpha $&$\beta_w$&$\phi_w$\\
		\hline
		\hline
	 \multirow{4}{*}{$\Xi_b^0 \to $} & \multirow{2}{*}{$\Sigma^+ K^ -$}&$ 7.52 \pm 1.29 \pm 0.27 $&$ 0.40 \pm 0.01 $&$ -0.04 \pm 0.02 $&$ 0.15 \pm 0.00 $\\
		&&$ 0.06 \pm 0.00 $&$ 0.18 \pm  0.00 $&$ -0.38 \pm 0.01 $&$ 0.76 \pm 0.01 $\\\cline{3-6}
		& \multirow{2}{*}{$\Sigma^+ \pi^ -$}&$ 6.62 \pm 1.09 \pm 0.86 $&$- 0.93 \pm 0.01 $&$ 0.08 \pm 0.04 $&$ 0.08 \pm 0.00 $\\
		&&$ -0.05 \pm 0.00 $&$ -0.01 \pm  0.00 $&$ 0.25 \pm 0.02 $&$ -2.88 \pm 0.02 $\\
		\hline	
		\multirow{8}{*}{$\Xi_b^- \to $}&\multirow{2}{*}{$\Lambda^0 K^-$}& $ 1.10 \pm 0.27 \pm 0.04 $&$ 0.41 \pm 0.01 $&$ -0.04 \pm 0.02 $&$ 0.15 \pm 0.00 $\\
		&&$ 0.06 \pm 0.00 $&$ 0.18 \pm  0.00 $&$ -0.39 \pm 0.01 $&$ 0.76 \pm 0.01 $\\\cline{3-6}
		&\multirow{2}{*}{$\Sigma^0 K^-$}		&$ 4.00 \pm 0.68 \pm 0.14 $&$ 0.40 \pm 0.01 $&$- 0.04 \pm 0.02 $&$ 0.15 \pm 0.00 $\\
		&&$ 0.06 \pm 0.00 $&$ 0.18 \pm  0.00 $&$ -0.38 \pm 0.01 $&$ 0.76 \pm 0.01 $\\\cline{3-6}
		&\multirow{2}{*}{$\Lambda^0 \pi^-$} & $ 0.98 \pm 0.24 \pm 0.13 $&$ -0.93 \pm 0.01 $&$ 0.08 \pm 0.04 $&$ 0.08 \pm 0.00 $\\
		&&$ -0.05 \pm 0.00 $&$- 0.01 \pm  0.00 $&$ 0.25 \pm 0.02 $&$ -2.88 \pm 0.02 $\\\cline{3-6}
		&\multirow{2}{*}{$\Sigma^0 \pi^-$}		&$ 3.52 \pm 0.58 \pm 0.46 $&$ -0.93 \pm 0.01 $&$ 0.08 \pm 0.00 $&$ 0.08 \pm 0.00 $\\
		&&$ -0.05 \pm 0.00 $&$ -0.01 \pm  0.00 $&$ 0.25 \pm 0.02 $&$ -2.88 \pm 0.02 $\\
		\hline
		\multirow{4}{*}{$\Lambda_b^0 \to $} & \multirow{2}{*}{$p K^ -$}&$ 5.52 \pm 0.65 \pm 0.19 $&$ 0.37 \pm 0.01 $&$ -0.04 \pm 0.02 $&$ 0.15 \pm 0.00 $\\
		&&$ 0.06 \pm 0.00 $&$ 0.18 \pm  0.00 $&$ -0.35 \pm 0.01 $&$ 0.76 \pm 0.01 $\\\cline{3-6}
		& \multirow{2}{*}{$p \pi^ -$}&$ 4.62 \pm 0.52 \pm 0.60 $&$- 0.91 \pm 0.006 $&$ 0.09 \pm 0.04 $&$ 0.08 \pm 0.00 $\\
		&&$ -0.05 \pm 0.00 $&$ -0.03 \pm  0.00 $&$ 0.25 \pm 0.02 $&$-2.88 \pm 0.02 $\\
		\hline
	\end{tabular}
	\end{center}
\end{table}
 In particular, the T violating asymmetries are found to be
 \begin{eqnarray}
 && \beta_w(\Xi_b^0 \to \Sigma^+ K^ -\,, \Xi_b^- \to \Lambda^0 K^-\,,  \Xi_b^- \to \Sigma^0 K^-\,, \Lambda_b^0 \to p K^ -)
 \nonumber\\\label{result1}
 &=&
 - 0.38 \pm 0.01\,,  - 0.39 \pm 0.01\,,   -0.38 \pm 0.01\,,   -0.35 \pm 0.01\,, \\
  && \beta_w(\Xi_b^0 \to \Sigma^+ \pi^ -\,, \Xi_b^- \to \Lambda^0 \pi^-\,,  \Xi_b^- \to \Sigma^0 \pi^-\,, \Lambda_b^0 \to p \pi^ -)
  \nonumber\\\label{result2}
  &=&
  0.25 \pm 0.02\,,   0.25 \pm 0.02\,,   0.25 \pm 0.01\,,   0.25 \pm 0.02\,, 
 \end{eqnarray}
 for $b\to d$ and $b\to s$ transitions in the standard model, respectively.
  As the absolute values of the asymmetries can be as large as $40\%$ and $25\%$
   , they can be measured in the experiments at LHCb,
  which provide a great opportunity to directly observe  TV. 
The main differences compared to the naive estimations in Eqs.~(\ref{newadd1}) and (\ref{newadd2}) are caused by the corrections from the effective Wilson coefficients, given by $|a_6/c_6| \approx 1.5$~\cite{GeneralModifiedBagModel,GeneralGeng:2021nkl,GeneralGeng:2016kjv,GeneralHsiao:2017tif,EffectiveWilson}.  Note that the results of $\alpha$ and $\beta$ have small uncertainties due to the correlations between $f_s$ and $g_a$, {\it i.e.} their ratios depend little on the model parameters. Nonetheless, it must be emphasized that the uncertainties from  the nonfactorizable effects
are not considered as they  can not be reliably calculated. However, the $\Xi_b^-$ decays are free from the W-exchange diagrams, making the results reliable.

\section{Cascade decay angular distributions}
We now concentrate on  the effects of the T-odd spin correlations. 
Although the cascade decays of ${\bf B}_n$ conserve  T-symmetry,  
the T-odd spin correlations in ${\bf B}_b \to {\bf B}_n P$ are handed down to the angular distributions of the sequential particles. 
For  the  level-2 cascade decays of
$\Xi_b^- \to \Lambda (\to p \pi ^ - )  K^- / \pi ^-$ and $\Xi_b^0 \to \Sigma^+(\to p \pi^0 ) K^- / \pi ^- $, 
 the angular distributions are given by
\begin{eqnarray}\label{level2}
&&{\cal D} _2 ( \vec{\Omega}) = \frac{1}{\Gamma}\frac{\partial^3 \Gamma }{\partial \cos \theta \partial \cos\theta_1 \partial \phi }=
\nonumber\\
&&\frac{1}{8\pi} \left[1 +P_b \alpha \cos \theta \cos \theta_1   + \alpha_b (\alpha \cos \theta_1 + P_b \cos \theta)+P_b \alpha(\beta \sin \phi 
-\gamma' \cos \phi 
)  \sin \theta \sin \theta_1\right],\,
\end{eqnarray}
where $P_b$ describe the $\Xi_b$ polarization fractions, 
$\alpha (= \alpha_{\Lambda ,\Sigma^+})$  represent the up-down asymmetry parameters in $\Lambda\to p \pi^-$ and $\Sigma^+ \to p \pi ^0$, and
\begin{equation}
\gamma' \equiv \frac{|{\cal S}|^2 - |{\cal P}|^2}{|{\cal S}|^2 + |{\cal P}|^2}\,.
\end{equation}
Here, $\theta$ and $\phi$ stand for the polar and azimuthal angles in the level-1 decays, 
while $\theta_1$ is the polar angle in the level-2 decays~(see FIGs.~2a and 2b, in which we have taken $P=K^-$ as an example).

\begin{figure}
	\begin{subfigure}{.48\textwidth}
		\centering
		\includegraphics[width=1\linewidth]{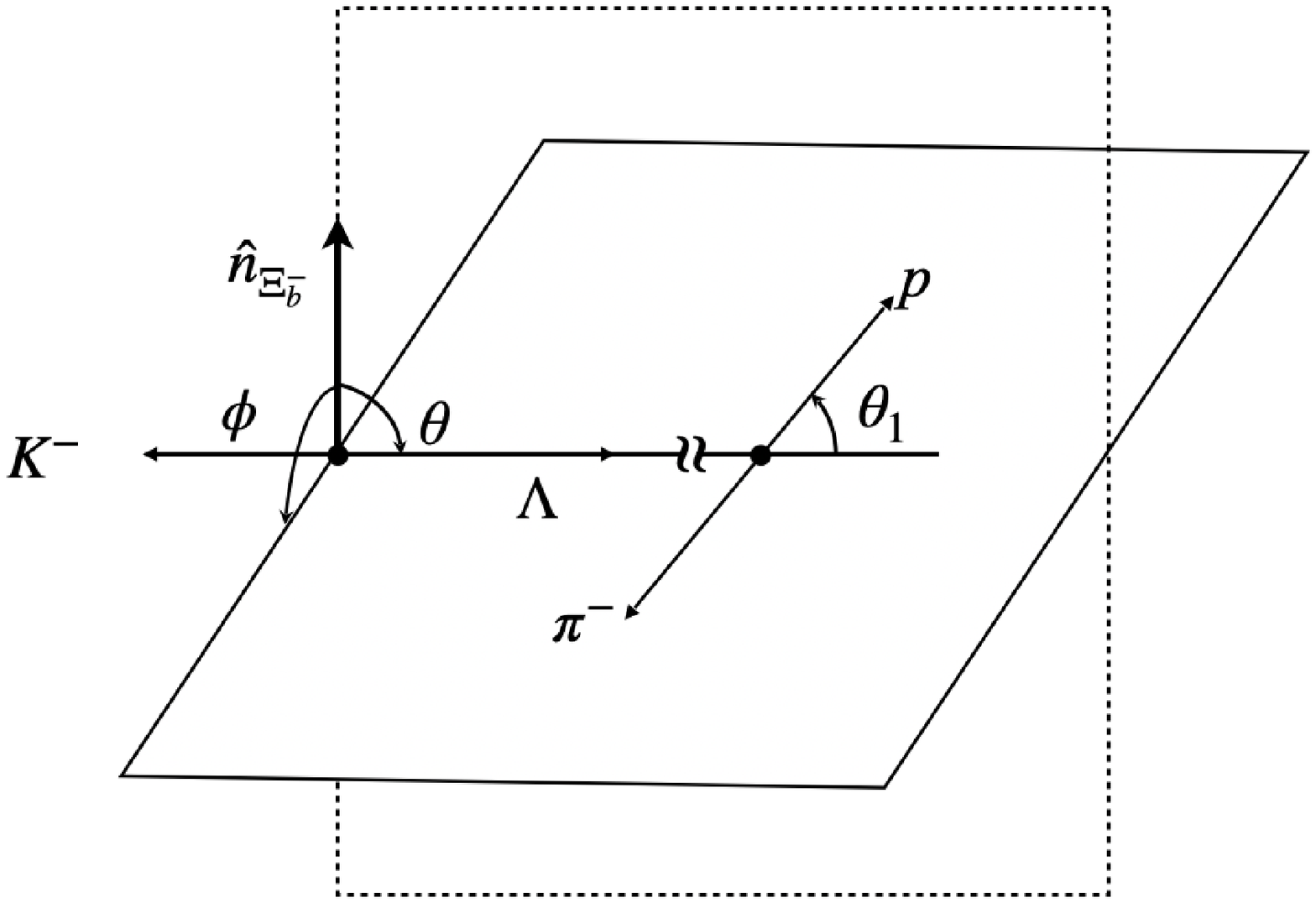}  
		\caption{ Cascade decays with $\Xi_b^-\to \Lambda(\to p \pi^-) K^-$.}
	\end{subfigure}
	\begin{subfigure}{.48\textwidth}
		\centering
		\includegraphics[width=1\linewidth]{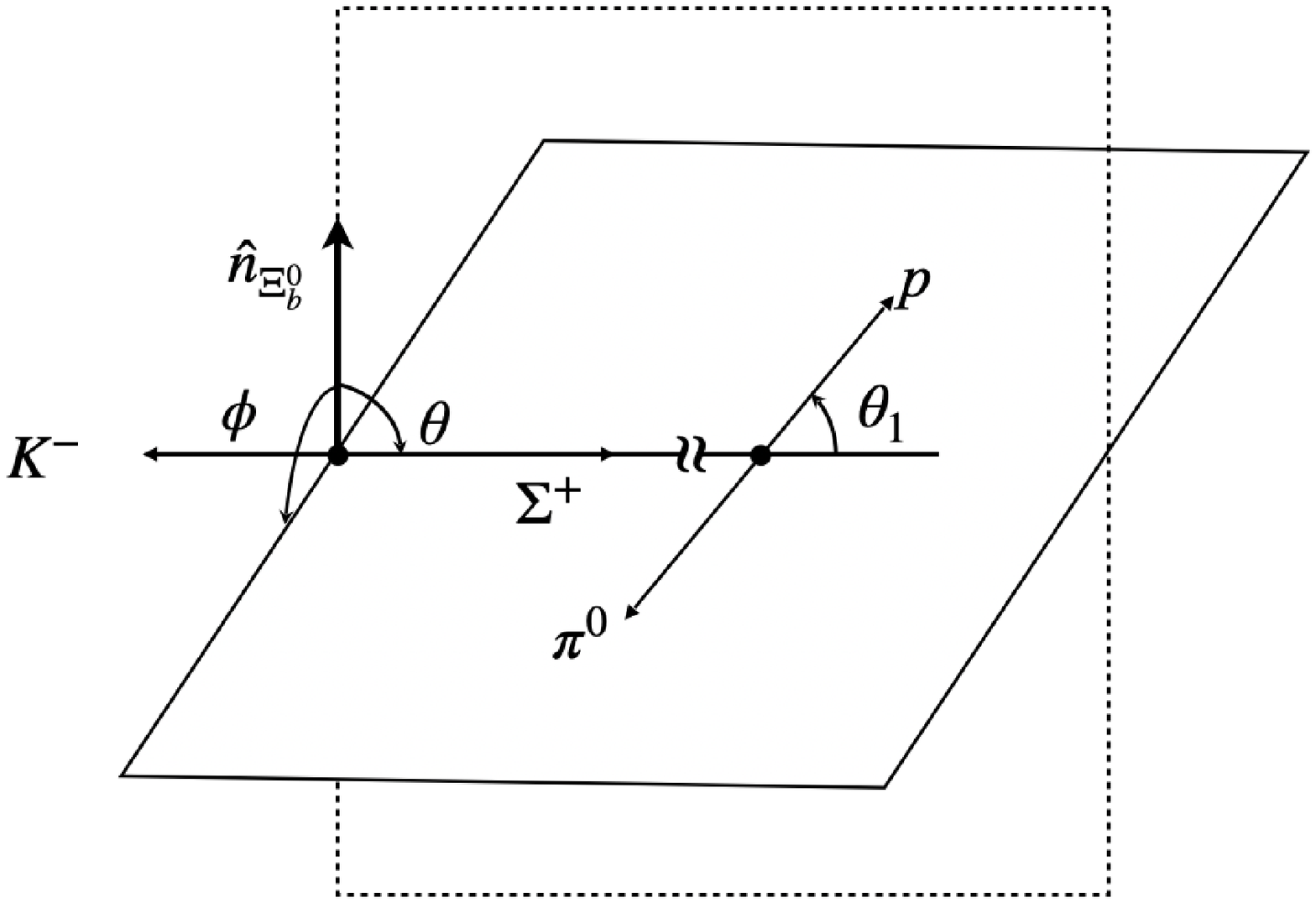}  
		\caption{Cascade decays with $\Xi_b^0\to \Sigma^+(\to p \pi^0) K^-$.}
	\end{subfigure}
	\newline
	
	\begin{subfigure}{.48\textwidth}
		\centering
		\includegraphics[width=1\linewidth]{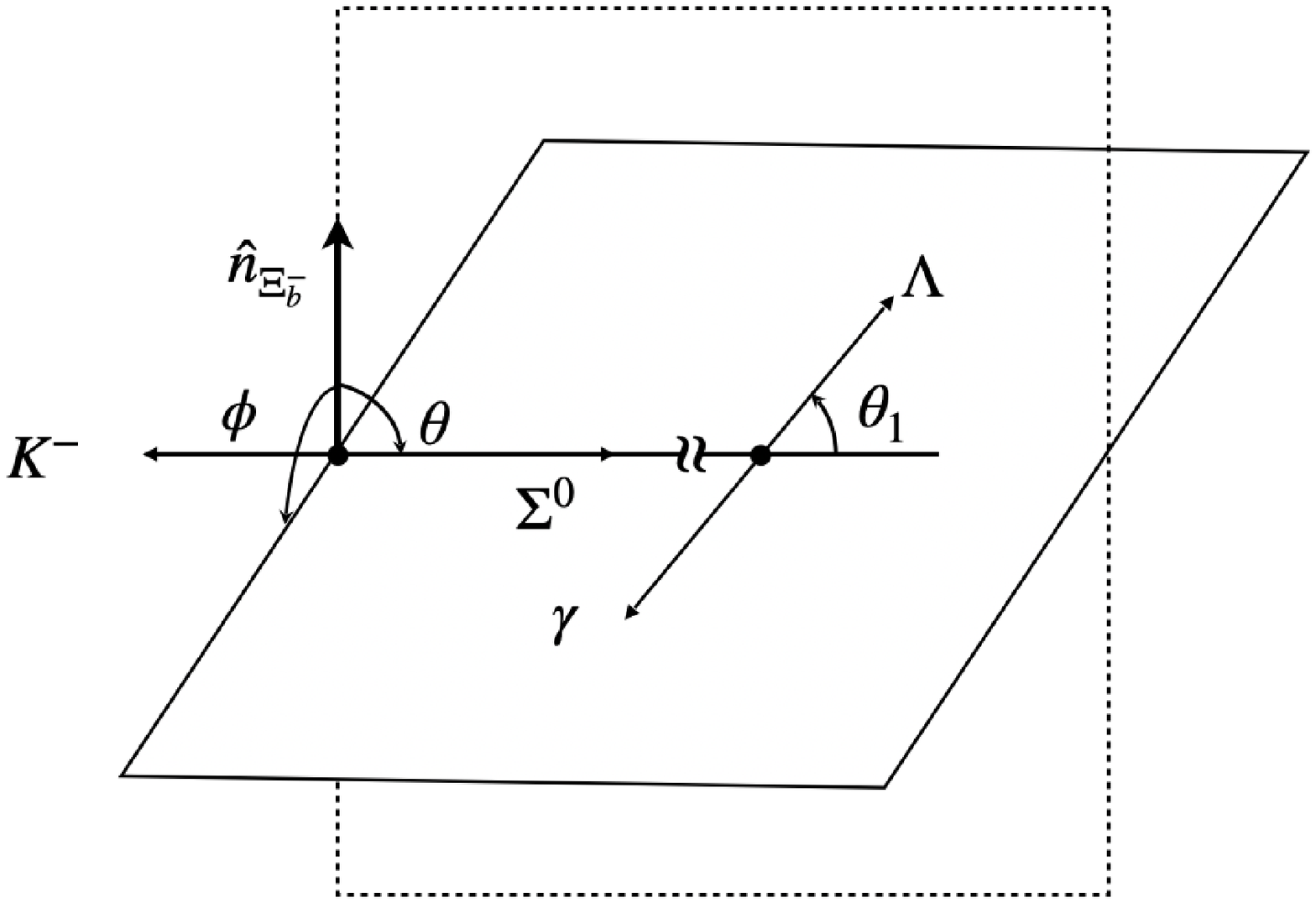}  
		\caption{Cascade decays with $\Xi_b^-\to \Sigma^0(\to \Lambda \gamma) K^-$.}
	\end{subfigure}
	\begin{subfigure}{.48\textwidth}
		\centering
		\includegraphics[width=1\linewidth]{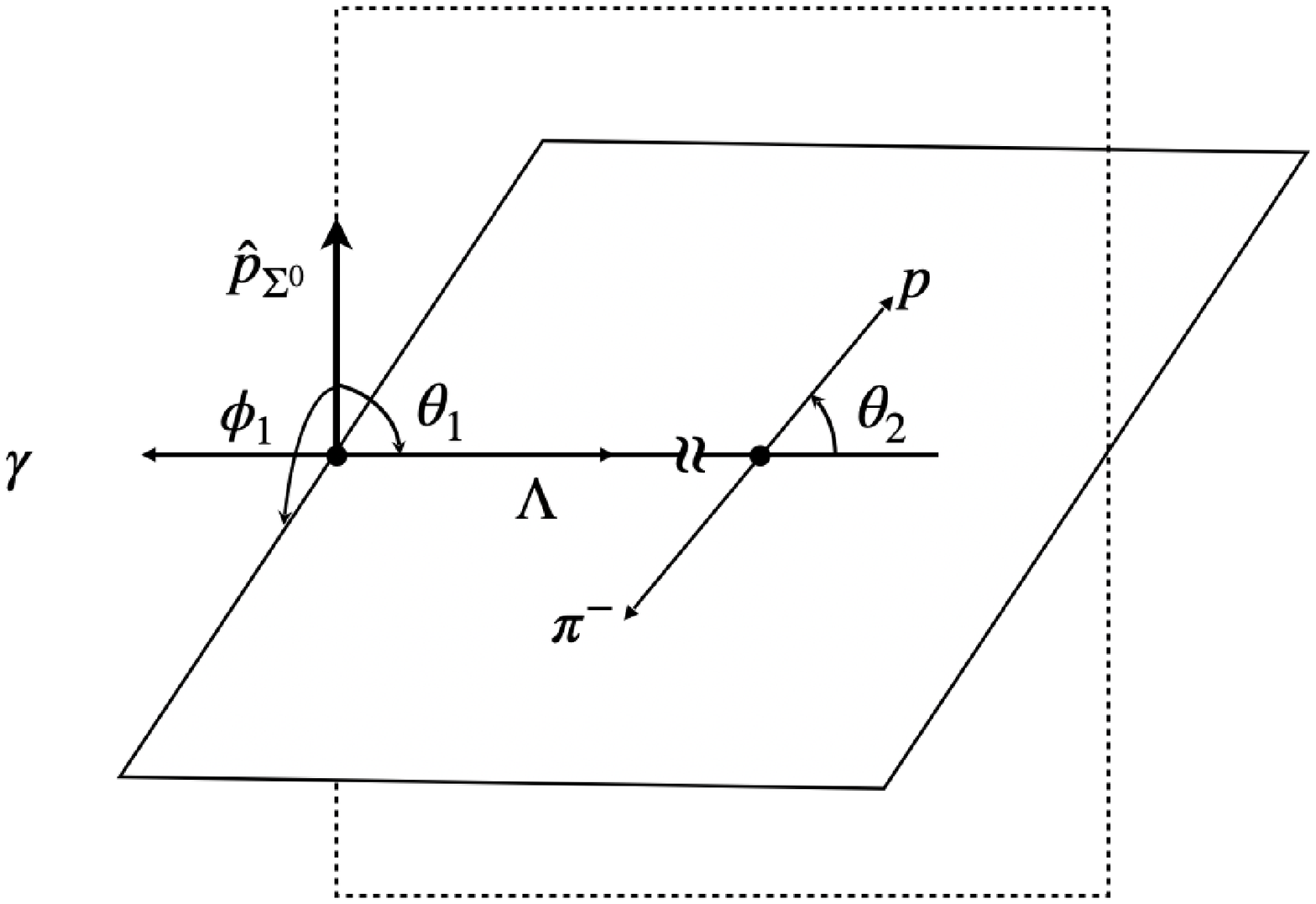}  
		\caption{Cascade decays with $\Sigma^0\to \Lambda(\to p \pi^-) \gamma$. }
	\end{subfigure}
	\centering 
	\caption{
		Angles in ${\cal D}(\vec{ \Omega})$ for cascade decays.
	}
\end{figure}

We see that the T-odd spin correlations indeed manifest themselves in ${\cal D}_2(\vec{ \Omega})$ and therefore can be observed in the experiments
at LHCb. 
	They are described by the asymmetries of  the dihedral angles between the initial and  cascade decay planes. To probe them, it is sufficient to consider the $\phi$-dependence only. 
	Based on the numbers of events,
	we define the naive T-odd observable as 
	\begin{equation}\label{simple}
	{\cal T}(N_+,N_-) = \frac{N_+ -  N_-}{N_+ + N_-}\,,
	\end{equation}
	where $N_\pm$ are the numbers of events with $\pi>\phi\ge0$ and $2\pi > \phi\ge\pi$, respectively. 
	For sufficient data points, we have
	\begin{equation}
	\lim\limits_{N_\pm\to \infty} {\cal T}(N+,N_-) =\frac{1}{\Gamma} \left(
	\int^\pi _ 0  \frac{\partial \Gamma}{\partial \phi} - 	\int^{2\pi} _\pi  \frac{\partial \Gamma}{\partial \phi}\right) 
	=\frac{\pi}{8}P_b \alpha \beta \,.
	\end{equation} 
Nonetheless, $\beta$ are always followed by $P_b$, requiring that $\Xi_b$ should be  polarized. 
This is a reasonable result since $\beta$ shall be able to  trace back to the spin correlations, and naturally demands that $P_b\neq 0$.

On the other hand, 
$\Xi_b^ - \to \Sigma^0( \to \Lambda ( \to p \pi^ - )  \gamma) K^- / \pi ^-$ is  a  level-3 cascade decay with the angular distribution,  given by
\begin{eqnarray}\label{level3}
{\cal D}_3 (\vec{\Omega})  &=& \frac{1}{\Gamma} \frac{\partial^5 \Gamma }{\partial \cos \theta \partial \cos\theta_1 \partial \cos \theta_2  \partial \phi  \partial \phi_1 }
\nonumber\\
&=&\frac{1}{32\pi^2}\left[1 -P_b \alpha_\Lambda \cos\theta \cos \theta_1 \cos \theta_2 +\alpha_b(P_b \cos \theta - \alpha_\Lambda \cos \theta_1 \cos \theta_2) \right.\nonumber\\
&&\left.+ P_b \alpha_\Lambda ( \gamma' \cos \phi - \beta \sin \phi ) \sin \theta \sin \theta_1 \cos \theta_2 \right] \,,
\end{eqnarray}
where the definitions of $\theta,\,\theta_1$ and $\phi$ are same as the previous ones, while 
$\phi_1~(\theta_2)$ stands for the  polar~(azimuthal) angle in the level-2~(3) decays, shown in FIGs.~2c and 2d. 
Due to the angular momentum conservation, 
${\cal D} _3(\vec{ \Omega})$ is independent of $\phi_1$.

Similar to ${\cal D}_2(\vec{ \Omega} ),$ $\beta$ are also followed by $P_b$ in ${\cal D}_3(\vec{\Omega})$, which is a sensible result. Notably, to observe $\beta$, it is necessary to include the  cascade  decays of $\Lambda$. To see this,  by  integrating ${\cal D}_3(\vec{ \Omega})$ over $\phi_1$ and $\cos \theta_2$,  
we find that  
\begin{equation}
\int^{2\pi}_0 \int^1_{-1}   {\cal D}_3(\vec{\Omega}) d\cos \theta_2 d \phi_1= 
\frac{1}{8\pi} \left(
1 + \alpha_b P_b \cos \theta
\right)\,,
\end{equation}
which is identical to ${\cal D}_2(\Omega)$ by setting $\alpha=0$. Due to this reason, as opposed to ${\cal D}_2(\vec{ \Omega})$, it is {\it not} sufficient to merely consider the $\phi$-dependence for observing TVEs, and there is no simple T violating observable as the one defined in Eq.~\eqref{simple}.

Note that  $\alpha_{\Lambda , \Sigma^+}$ can be determined by the experiments in the hyperon decays, given by~\cite{pdg,BESIII:2018cnd}
\begin{equation}
\alpha_\Lambda = 0.731 \pm 0.014\,,\,\,\,\,\,\,\,\alpha_{\Sigma^+} = -0.982\pm 0.014\,.
\end{equation}
However, the values of $P_b$  depend on the production method of the bottom baryons, which are  ${\cal O}(10\%)$
at LHCb~\cite{ExpPolarized3, Hiller:2007ur}.
Theoretically, it has been   suggested that $P_b$ can be as large as 20\%~\cite{French0}.
Their actual values could be measured  from the decays with enough data points~\cite{BESIII:2017kqg,BESIII:2019odb}. 
For instances,  we can obtain $P_b$  from the angular distributions of $\Xi_b^{0,+} \to \Xi^{0,+} J/\psi(\to l^+ l^-)$,  similar to  $\Lambda_b\to \Lambda  J/\psi(\to l^+ l^-)$~\cite{ExpPolarized1,ExpPolarized2,ExpPolarized3},
in which $P_b$ is found to be
	$[-0.06, 0.05]$, $[-0.04, 0.05]$ and $[-0.01, 0.07]$ at the central energies of 7, 8 and 13~TeV of proton-proton collisions, respectively, with 68\% credibility level intervals~\cite{ExpPolarized3}.
Finally, we note that the branching ratios of $\Xi_b^{0,+} \to \Xi^{0,+} J/\psi$ are found to be  around  $10^{-4}$ in terms of the $SU(3)$ flavor symmetry with the experimental input of ${\cal B}(\Lambda_b\to \Lambda  J/\psi) = (5.8 \pm 0.8 )\times 10^{-5}$~\cite{pdg}. A large decay branching ratio
can enhance
 the experiments to extract $P_b$ with high precision. 

\section{Conclusions}
Aiming on probing TV, we have analyzed  the most simple cases in
the charmless two-body decays of  ${\bf B}_b \to {\bf B}_n M$ with $M= (P,V,\gamma)$,   in which the TVEs are described by the T-odd spin correlations.
We have found that the T violating parameters of 
$ \beta_w(\Xi_b^{0,-} \to \Sigma^{+,0} K^ -\,, \Xi_b^{-,0} \to (\Lambda^0,p) K^-)$
and 
$ \beta_w(\Xi_b^{0,-} \to \Sigma^{+,0} \pi^ -\,, \Xi_b^{-,0} \to (\Lambda^0,p) \pi^-)$
are around $-40\%$ and $25\%$ with small uncertainties in  the standard model, 
which can be measured by the experiments at LHCb, respectively.
We emphasize that in contrast to other CP asymmetries, the signs of $\beta_w$ are insensitive to the strong phases which can not be reliably calculated. Hence, if the experiments obtain opposite signs of $\beta_w$,
it would be a smoking gun for new physics, such as the right-handed current beyond the standard model.


We have studied the angular distributions of the $\Xi_b$ cascade decays and identified the TVEs, in which polarized $\Xi_b$ are needed
in order to observe TV. 
We have also shown that
  $P_b$ can be obtained from  the angular distributions of  $\Xi_b \to \Xi J/\psi(\to l^+ l^-)$ 
   in the experiments at LHCb since their decay branching ratios are 
   predicted to be around $10^{-4}$ by virtual of  the $SU(3)$ flavor symmetry.


\begin{thebibliography}{99}
\bibitem{footnote1}
The decay branching ratios in this work are slightly different with those in Ref.~\cite{GeneralModifiedBagModel} as we have updated
the experimental inputs of the CKM matrix elements.
\end{thebibliography}
\end{document}